\documentclass[useAMS,usenatbib,usegraphicx]{mn2e}

\newcommand{\asec}{$^{\prime\prime}$}

\def\HNC{HN$^{13}$C}
\def\H{N$_{2}$H$^{+}$}
\def\D{N$_{2}$D$^{+}$}

\def\AMM{NH$_3$}

\def\HCO{\mbox{HCO$^+$}}

\def\kms{\mbox{km~s$^{-1}$}}
\def\cmc{cm$^{-3}$}

\def\Vlsr{$V_{\rm LSR}$}
\def\Dfrac{$D_{\rm frac}$}

\def\Lsun{$L_{}$}
\def\Tex{\mbox{$T_{\rm ex}$}}

\def\Tk{\mbox{$T_{\rm k}$}}

\def\Tmb{\mbox{$T_{\rm MB}$}}
\def\Ta{\mbox{$T_{\rm a}^*$}}

\def\etamb{$\eta_{\rm MB}$}

\title[DNC/HNC ratio in high-mass star forming cores]{DNC/HNC and \D/\H\ ratios in high-mass star forming cores}
\author[F. Fontani et al.]{F. Fontani$^{1}$\thanks{E-mail:
fontani@arcetri.astro.it}, T. Sakai$^{2}$, K. Furuya$^{3}$, N. Sakai$^{4}$, Y. Aikawa$^{3}$, S. Yamamoto$^{4}$
\\
\\
$^{1}$INAF-Osservatorio Astrofisico di Arcetri, L.go E. Fermi 5, Firenze, I-50125, Italy \\
$^{2}$Graduate School of Informatics and Engineering, The University of Electro-Communications, Chofu, Tokyo 182-8585, Japan \\
$^{3}$Department of Earth and Planetary Sciences, Kobe University, Kobe 657-8501, Japan \\
$^{4}$Department of Physics, Graduate School of Science, The University of Tokyo, Tokyo 113-0033, Japan \\
}
\begin{document}

\date{Accepted date. Received date; in original form date}

\pagerange{\pageref{firstpage}--\pageref{lastpage}} \pubyear{2011}

\maketitle

\label{firstpage}

\begin{abstract}
 Chemical models predict that the deuterated fraction (the column density ratio between a molecule 
containing D and its counterpart containing H) of \H , \Dfrac (\H ), high in massive pre-protostellar cores, 
is expected to rapidly drop of an order of magnitude after the protostar birth, while that of HNC, 
\Dfrac (HNC), remains constant for much longer. We tested these predictions by deriving \Dfrac (HNC) 
in 22 high-mass star forming cores divided in three different evolutionary stages, from high-mass 
starless core candidates (HMSCs, 8) to high-mass protostellar objects (HMPOs, 7) to 
Ultracompact HII regions (UCHIIs, 7). For all of them, \Dfrac (\H ) was already determined 
through IRAM-30m Telescope observations, which confirmed the theoretical rapid decrease of 
\Dfrac(\H ) after protostar birth (Fontani et al. 2011). Therefore our comparative study is not affected 
by biases introduced by the source selection. We have found average \Dfrac (HNC) of 0.012, 0.009 
and 0.008 in HMSCs, HMPOs and UCHIIs, respectively, with no statistically significant differences 
among the three evolutionary groups. These findings confirm the predictions of the chemical 
models, and indicate that large values of \Dfrac (\H ) are more suitable than large values of 
\Dfrac (HNC) to identify cores on the verge of forming high-mass stars, likewise what found 
in the low-mass regime.
\end{abstract}


\section{Introduction}
\label{intro}

The process of deuterium enrichment in molecules from HD, the
main reservoir of deuterium in molecular clouds, is initiated by three 
exothermic ion-molecule reactions (e.g.~Millar et al.~1989):
\begin{equation}
{\rm H_3^+ + HD \rightarrow H_2D^+ + H_2}\;,
\label{eqa}
\end{equation}
\begin{equation}
{\rm CH_3^+ + HD \rightarrow CH_2D^+ + H_2}\;,
\label{eqb}
\end{equation}
\begin{equation}
{\rm C_2H_2^+ + HD \rightarrow C_2HD^+ + H_2} \;.
\label{eqc}
\end{equation}
Since the backward reactions are endothermic by 232,
390 and 550 K, respectively, in cold environments
(e.g.~$T_{\rm kin} \leq 20$ K for reaction (1))
they proceed very slowly, favouring the formation of deuterated ions.
Moreover, the freeze-out of CO and other neutrals, particularly relevant
in high-density gas ($n_{\rm H_2}\geq 10^4$ \cmc ), further boosts the 
deuteration process (e.g.~Bacmann et al.~\citeyear{bacmann03},
Crapsi et al.~\citeyear{crapsi05}, Gerin et al.~\citeyear{gerin06}). 
Therefore, in dense and cold cores the deuterated fraction, \Dfrac , 
defined as the abundance ratio between a deuterated molecule and its 
hydrogenated counterpart, is expected to be much higher than the
average [D/H] interstellar abundance (of the order of $10^{-5}$,
Oliveira et al.~\citeyear{oliveira03}, Linsky et al. ~\citeyear{linsky06}).
Because of the changements in physical and chemical properties of a star
forming core, its \Dfrac\ is expected to change with the evolution too. 
Specifically, \Dfrac\ is predicted to increase when a pre--protostellar core 
evolves towards the onset of gravitational collapse, as the core 
density profile becomes more and more centrally peaked (due to the 
temperature decrease at core centre, e.g.~Crapsi et al.~\citeyear{crapsi07}), 
and then it drops when the young stellar object formed 
at the core centre begins to heat its surroundings (see e.g. Caselli et al.~\citeyear{caselli02}).

While this net drop in \Dfrac\ before and after the protostellar birth in 
\Dfrac (\H ) is clearly observed in both low-mass (Crapsi et al.~\citeyear{crapsi05}, 
Emprechtinger et al.~\citeyear{emprechtinger09}) and high-mass
(Fontani et al.~\citeyear{fontani11}, Chen et al.~\citeyear{chen11}) star-forming
cores, other species show deviations from this general scenario. 
For example, DNC is produced in the gas from the same route reaction
as \D , namely reaction (\ref{eqa}), so that \Dfrac (HNC) and \Dfrac (\H ) 
are expected to vary similarly with temperature (Turner~\citeyear{turner}).
However, Sakai et al.~(\citeyear{sakai12}) have measured \Dfrac (HNC)
in a sample of 18 massive cores including both infrared-dark starless
cores and cores harbouring high-mass protostellar objects, and found that 
\Dfrac (HNC) in the starless cores is only marginally higher than
that measured in the protostellar cores. 
This 'anomaly' could be explained by the fact that the destruction processes 
of \D\ are much faster than those of DNC: being an ion, \D\ can recombine 
quickly (few years) with CO and/or electrons, while the neutral DNC has 
to be destroyed by ions (such as \HCO\ and/or H$_3^+$) through much 
slower ($10^4 - 10^5$ yrs) chemical reactions (Sakai et al.~\citeyear{sakai12}).

The chemical models of Sakai et al.~(\citeyear{sakai12}), are able to 
partially reproduce the observational
results obtained by Fontani et al.~(\citeyear{fontani11}) and Sakai et
al.~(\citeyear{sakai12}).
We have performed chemical calculations similar to those in Sakai et al.~(\citeyear{sakai12}), 
and obtained the consistent results (see Section 4.2 for the details of our chemical model): 
the \D /\H\ ratio approaches $\sim 0.1$ during the 
cold pre--protostellar phase and drops quickly to $\sim 0.01$ after the 
protostellar birth because very sensitive
to a temperature growth (Fig.~1, panel (a)), while the DNC/HNC ratio 
remains relatively high (above $\sim 0.01$) even after a rapid temperature rise, 
and decreases in timescales of several 10$^4$ yrs. On the other hand, the \D /\H\ 
drops much more quickly, in less than 100 yrs (Fig.~1, panel (b)).
However, the different criteria adopted by Fontani et al.~(\citeyear{fontani11})
and Sakai et al.~(\citeyear{sakai12}) to select the targets does not allow
for a consistent observational comparison between the two deuterated fractions,
as well as between models and data.
%

\begin{figure*}
\begin{minipage}{160mm}
 \begin{center}
 \resizebox{\hsize}{!}{\includegraphics[angle=0]{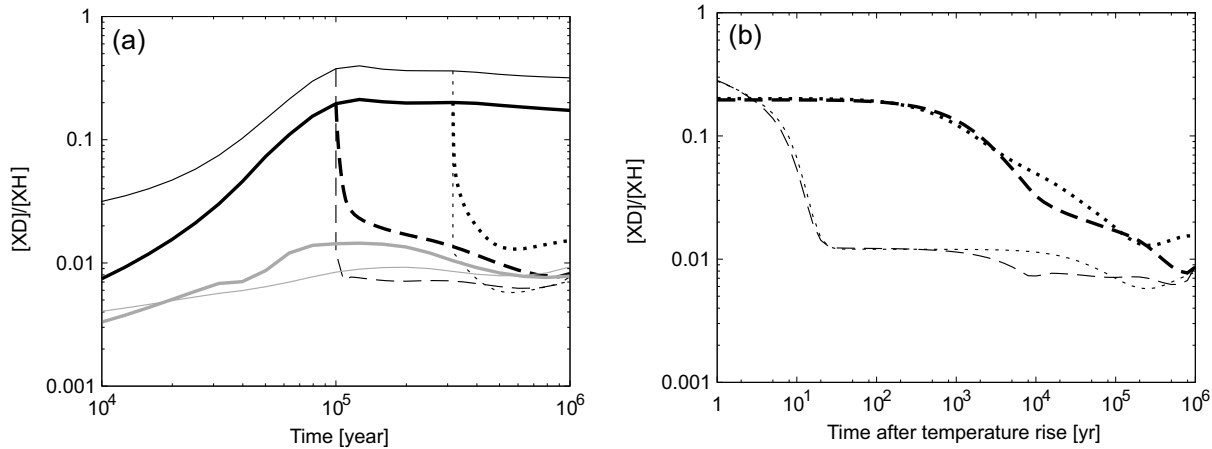}}
 \caption[]
 {\label{models}{{\bf (a)}: chemical model calculation of the time dependence of DNC/HNC (thick
lines) and N$_2$D$^+$/N$_2$H$^+$ (thin lines).  In addition to the constant temperature cases
(10 K = solid black, 30 K = solid grey), the cases in which the temperature suddenly rises
from 10 to 30 K at a given age, 1$\times$10$^5$ yr (dashed) or 3$\times$10$^5$ yr (dotted)
are shown. See Section~\ref{model} for the detail of the chemical model.
\newline
{\bf (b)}: same as panel {\bf (a)}, but the time starts from when the
temperature increases from 10 K to 30 K.
}}
 \end{center}
 \end{minipage}
\end{figure*}

In this paper we report observations performed with the Nobeyama-45m Telescope
in the DNC and HN$^{13}$C (1--0) rotational transitions towards 22 high-mass
cores harbouring different stages of the high-mass star formation process, 
in which \Dfrac (\H ) was already measured through
observations of the IRAM-30 Telescope (Fontani et al.~\citeyear{fontani11}). 
In this way, our study is not affected by observational biases possibly introduced by 
the source selection. 
The main aim of the work is to test in the same
sample of objects whether the \D /\H\ and DNC/HNC ratios trace 
differently the thermal history of high-mass cores despite the similar chemical origin,
as predicted by Sakai et al.~(\citeyear{sakai12}). The sample, selected as
explained in Fontani et al.~(\citeyear{fontani11}), is divided in 8 high-mass 
starless cores (HMSCs), 7 high-mass protostellar objects (HMPOs) and 
7 ultracompact HII regions (UC HIIs), so that all main evolutionary groups 
of the high-mass star formation process are almost equally represented. 
We stress that all HMSCs, except I22134--B, have been previously
classified as 'quiescent' by Fontani et al.~(\citeyear{fontani11}) to distinguish
them from 'perturbed' cores, in which external phenomena (passage of outflows, 
shocks, nearby infrared objects) can have affected significantly the physical-chemical
properties of the gas, as discussed in Fontani et al.~(\citeyear{fontani11}).

In Sect.~\ref{obs} we give an overview of the technical details of the observations;
Sect.~\ref{res} presents the main observational results, which are discussed in 
Sect.~\ref{dis}, including a detailed comparison with chemical models. A summary 
of the main findings of the paper are given in Sect.~\ref{con}.

\section{Observations}
\label{obs}

The \HNC (1--0) and DNC(1--0) transitions were observed with the NRO 45 m 
Telescope in may 2012 towards 22 out of the 27 cores already observed by Fontani
et al.~(\citeyear{fontani11}) in \D (2--1) and \H (3--2). The source coordinates,
as well as some basic properties of the star forming regions where they
are embedded (LSR velocity of the parental core, distance to the Sun, 
bolometric luminosity, \Dfrac (\H ) as measured by Fontani et al.~\citeyear{fontani11}) 
are listed in Table~\ref{tab_sou}. 
The two transitions were observed simultaneously by using the 
sideband-separating superconductor-insulator-superconductor receiver, T100 
(Nakajima et al.~\citeyear{nakajima08}).

Some important spectroscopic parameters of the lines observed and the 
main technical parameters are listed in Table~\ref{tab_lin}.
The half-power beam width is about 21\asec\ and 18\asec\ at 76 (DNC (1--0))
and 87 GHz (\HNC (1--0)), respectively, similar to the
beam width of the IRAM-30m Telescope at the frequency of the \D (2--1)
line ($\sim 15$ \asec , Fontani et al.~\citeyear{fontani11}).
The main beam efficiency ($\eta_{\rm MB}$) is 0.53 and 0.43 at 76 and 87 GHz, 
respectively. We derived the main beam temperature (\Tmb ) from the antenna temperature 
(\Ta ) by using the relation \Tmb\ = \Ta /$\eta_{\rm MB}$, where $\eta_{\rm MB}$
is the main beam efficiency (see Table~\ref{tab_lin}).
For all the observations, we used digital backends SAM45 
(bandwidth = 500~MHz, frequency resolution = 122.07~kHz). 
The telescope pointing was checked by observing nearby SiO maser source
every one to two hours, and was maintained to be better than 5\asec . 
The line intensities were calibrated by the chopper wheel method. 
All the observations were carried out with the position switching mode. 

\begin{table*}
\begin{center}
\begin{minipage}{140mm}
\caption[] {List of the observed sources. Col.~4, 5 and 6 show the velocity at which
we centred the spectra (corresponding to the systemic velocity), 
the source distance and bolometric luminosity of the associated star 
forming region, respectively. This latter is a very rough first approximation 
of the core luminosity because it is based on infrared measurements having 
poor angular resolution.
We adopt as source names those adopted by Fontani et al.~(\citeyear{fontani11}), 
who took them from the reference papers listed in Col.~7,.
For completeness, \Dfrac (\H ) derived by Fontani et al.~(\citeyear{fontani11}) for
each core is given in Col.~8.}
\label{tab_sou}
\normalsize
\begin{tabular}{lccccccc}
\hline \hline
source& RA(J2000) & Dec(J2000) & \Vlsr\ & $d$ & $L_{\rm bol}$ & Ref. & \Dfrac(\H ) \\
    & h m s & $o$ $\prime$  $\prime\prime$ & \kms\   & kpc & \Lsun\ & &  \\
\cline{1-8}
\multicolumn{8}{c}{HMSC} \\
\cline{1-8}
I00117-MM2 $^{a}$ & 00:14:26.3	& +64:28:28 & $-36.3$ & 1.8  & $10^{3.1}$ & (1) & 0.32  \\
G034-G2(MM2) $^{a}$ & 18:56:50.0 & +01:23:08 &  $+43.6$ & 2.9 & $10^{1.6}$ $^{r}$ & (4) & 0.7 \\
G034-F2(MM7) $^{a}$ & 18:53:19.1  & +01:26:53 & $+57.7$  & 3.7 & $10^{1.9}$ $^{r}$   & (4) & 0.43 \\
G034-F1(MM8) $^{a}$ & 18:53:16.5  & +01:26:10 & $+57.7$  & 3.7 & -- & (4) & 0.4 \\
G028-C1(MM9)  $^{a}$ & 18:42:46.9  & $-$04:04:08 & $+78.3$  & 5.0  & -- & (4) & 0.38 \\
I20293-WC $^{a}$ & 20:31:10.7  &	 +40:03:28 & $+6.3$ & 2.0 & $10^{3.6}$ & (5,6) & 0.19 \\
I22134-G $^{b}$  $^{w}$  & 22:15:10.5  &   +58:48:59 & $-18.3$  & 2.6 & $10^{4.1}$ & (7) & 0.023 \\
I22134-B $^{b}$  & 22:15:05.8 &	+58:48:59 & $-18.3$ & 2.6  & $10^{4.1}$ & (7) & 0.09 \\
\cline{1-8}
\multicolumn{8}{c}{HMPO}   \\
\cline{1-8}
I00117-MM1 $^{a}$ & 00:14:26.1	& +64:28:44 & $-36.3$  & 1.8 & $10^{3.1}$ & (1) & $\leq 0.04$\\
18089--1732 $^{b}$  & 18:11:51.4 & $-$17:31:28 & $+32.7$  & 3.6 & $10^{4.5}$ & (9) & 0.031 \\
18517+0437 $^{b}$  & 18:54:14.2 & +04:41:41 & $+43.7$  & 2.9 & $10^{4.1}$ & (10) & 0.026 \\
G75-core $^{a}$ & 20:21:44.0 &	+37:26:38 & $+0.2$  & 3.8 & $10^{4.8}$ & (11,12) & $\leq 0.02$ \\
I20293-MM1 $^{a}$ & 20:31:12.8 &	 +40:03:23 & $+6.3$ & 2.0 & $10^{3.6}$ & (5) & 0.07 \\
I21307 $^{a}$ & 21:32:30.6  &    +51:02:16  & $-46.7$ & 3.2 & $10^{3.6}$ & (13) & $\leq 0.03$\\ 
I23385 $^{a}$ & 23:40:54.5 &      +61:10:28 & $-50.5$   & 4.9 & $10^{4.2}$ & (14) & 0.028 \\
\cline{1-8}
\multicolumn{8}{c}{UC HII}   \\
\cline{1-8}
G5.89--0.39 $^{b}$  & 18:00:30.5  & $-$24:04:01 & $+9.0$ & 1.28 & $10^{5.1}$ & (15,16) & 0.018\\
I19035-VLA1 $^{b}$  & 19:06:01.5 &	+06:46:35 & $+32.4$  & 2.2 & $10^{3.9}$ & (11) & 0.04 \\
19410+2336 $^{a}$ & 19:43:11.4 &    +23:44:06 & $+22.4$ & 2.1 & $10^{4.0}$ & (17) & 0.047 \\
ON1 $^{a}$ & 20:10:09.1  &     +31:31:36 & $+12.0$  & 2.5 & $10^{4.3}$ & (18,19) & 0.017 \\
I22134-VLA1 $^{a}$ & 22:15:09.2 &	+58:49:08 & $-18.3$ & 2.6 & $10^{4.1}$ & (11) & 0.08 \\
23033+5951 $^{a}$ & 23:05:24.6 & +60:08:09 & $-53.0$  & 3.5 & $10^{4.0}$ & (17) & 0.08 \\
NGC7538-IRS9 $^{a}$  & 23:14:01.8   &   +61:27:20 & $-57.0$  & 2.8 & $10^{4.6}$ & (8) & 0.030 \\
\hline
\end{tabular}

 $^{a}$ Observed in \H\ (3--2) and \D\ (2--1); \\
 $^{b}$ Observed in \H\ (1--0), \H\ (3--2), and \D\ (2--1); \\
 $^{c}$ Observed in \H\ (1--0) and \D\ (2--1); \\
 $^{w}$ "warm" HMSC; \\
 $^{r}$ Luminosity of the core and not of the whole associated star-forming region (Rathborne et al.~\citeyear{rathborne}); \\
 References: (1) Palau et al.~(\citeyear{palau10}); 
 (2) Busquet et al.~(\citeyear{busquet11}); 
 (3) Beuther et al.~(\citeyear{beuther07}); 
 (4) Butler \& Tan~(\citeyear{bet}); 
 (5) Palau et al.~(\citeyear{palau07}); 
 (6) Busquet et al.~(\citeyear{busquet}); 
 (7) Busquet~(\citeyear{busquetphd}); 
 (8) S\'anchez-Monge et al.~(\citeyear{sanchez}); 
 (9) Beuther et al.~(\citeyear{beuther04}); 
 (10) Schnee \& Carpenter~(\citeyear{schnee}); 
 (11) S\'anchez-Monge~(\citeyear{sanchez11}); 
 (12) Ando et al.~(\citeyear{ando}); 
 (13) Fontani et al.~(\citeyear{fonta04a}); 
 (14) Fontani et al.~(\citeyear{fonta04b}); 
 (15) Hunter et al.~(\citeyear{hunter}); 
 (16) Motogi et al.~(\citeyear{motogi}); 
 (17) Beuther et al.~(\citeyear{beuther02}); 
 (18) Su et al.~(\citeyear{su}); 
 (19) Nagayama et al.~(\citeyear{nagayama}).  
  \end{minipage}
  \end{center}
\end{table*}

\begin{table*}
\begin{center}
\caption[] {Observed transitions and technical parameters}
\label{tab_lin}
\begin{tabular}{lcccccccc}
\hline
Transition & Rest Frequency  &	$E_{\rm u}/k$ &  $\mu_0$ &	BW & $\Delta_\nu$ & HPBW &	\etamb\ &	$T_{\rm sys}$ \\
		& (GHz)            &	(K)	         &  (D)	&      (MHz)	&       (kHz)       &	(arcsec)	&	&     (K) \\
\hline
DNC	(1--0) &	76.305727 &	3.66	&     3.05	&    40            &    	37	&            21	&                  0.53 &	250 \\
HN$^{13}$C (1--0) & 87.090850 &	4.18	&     3.05	&    40	     &          37	&            18	&                  0.43 &	150 \\
\hline
\end{tabular}
\end{center}
\end{table*}

\section{Results}
\label{res}

\subsection{Detection rates and line profiles}
\label{line_profiles}

We have detected DNC (1--0) emission in: 6 out of 8 HMSCs, 3 out of 7 HMPOs, and
5 out of 7 UCHIIs. \HNC\ (1--0) has been detected towards all cores detected in DCN,
except I00117-MM2, undetected in \HNC\ but detected in DNC (although a faint line
at $\sim 2.5 \sigma$ rms level is possibly present in the spectrum).  
Therefore, the detection rate follows a trend with core evolution similar to the
one observed in \D\ by Fontani et al.~(\citeyear{fontani11}): its maximum is in the
HMSC phase, then it decreases in the HMPO phase but does not decrease further
in the later stage of UC HII region. In the Appendix--A, available online, we show all spectra
(Figs.~A.1 to A.6). 
They have been analysed with the CLASS program, which is part of the GILDAS 
software\footnote{The GILDAS software is available at http://www.iram.fr/IRAMFR/GILDAS}
developed at the IRAM and the Observatoire de Grenoble. 

Both transitions have a hyperfine structure due to electric
quadrupole interactions of nitrogen and deuterium nuclei.
We tried to fit the observed spectral line profiles by considering these
components through the command 'method hfs' into CLASS.
However, given that the maximum separation between the components 
is 1.28~\kms\ and 0.73~\kms\ for DNC and \HNC , respectively (van der Tak
et al.~\citeyear{vandertak}), i.e.~smaller than the typical velocity widths 
($\sim 1 -2$ \kms ), the method failed. 
Due to this, and because the observed spectra typically
show single-peaked profiles, the lines have been fitted with Gaussian 
functions. In Tables~\ref{tab_fit_dnc} and \ref{tab_fit_hnc} we give the 
main parameters of the lines derived through Gaussian fits: integrated
area ($A$), full width at half maximum ($\Delta v$), peak temperature
($T_{\rm pk}$) and 1$\sigma$ rms of the spectrum.

Asymmetric profiles and hints of non-Gaussian high-velocity wings are 
detected in the DNC(1--0) spectrum of G034--F1 and G034--F2 (Fig.~A.1), 
and in the \HNC (1--0) spectrum of G034--F2, G028--C1 (Figs.~A.1), 
18089--1732, 18517+0437 (Fig.~A.3), G5.89--0.39 
and I19035--VLA1 (Fig.~A.5), 23033+5951 and NGC7538--IRS9 (Fig.~A.6).
The non-Gaussian emission in the wings can be naturally attributed to outflows in the
HMPOs and UCHIIs, although neither DNC nor \HNC\ are typical
outflow tracers. On the other hand, this explanation is not plausible 
in the HMSCs G034--F1, G034--F2 and G028--C1, which do not
show any clear star formation activity. Other possibilities
could be the presence of multiple velocity components, or high
optical depth effects. High optical depths seem unlikely for the 
DNC and \HNC\ lines because these species are not very
abundant, and also because the major effect should be at line centre, 
while the asymmetries are seen at the edges of the lines (see
e.g. G034--F1 in~Fig.~A.1). 
The presence of multiple velocity components is 
the most realistic scenario, especially in the HMSCs. 
However, only with higher angular resolution observations one 
can shed light on the origin of these asymmetric line shapes.
 
\subsection{Line widths}
\label{line_widths}

In Fig.~\ref{line_widths} we compare the line widths of DNC(1--0) to both
those of the \HNC (1--0) and \D (2--1) transitions detected by 
Fontani et al.~(\citeyear{fontani11}). Globally, DNC and \HNC (1--0) have
comparable line widths (left panel in Fig.~\ref{line_widths}) regardless 
of the evolutionary stage of the cores.
This global trend confirms that the two transitions arise from gas with 
similar turbulence, as already found in the massive star-forming cores studied
by Sakai et al.~(\citeyear{sakai12}). 
Incidentally, we note that four HMSCs have \HNC (1--0) line widths  
in between $\sim 2$ and 3 \kms , i.e. almost twice the corresponding 
DNC(1--0) line widths. 
However, the low signal to noise ratio in the spectra,
and the fact that the kinematics in the targets may be complex
and due to the superimposition of several velocity components (as
suggested by the deviations from the Gaussian line shape, 
see Sect.~\ref{line_profiles}), 
could explain the different line widths of DNC and \HNC\ in these
sources.

A similar comparison between the DNC(1--0) line widths and those 
of the \D (2--1) transition is shown in the right panel of Fig.~\ref{line_widths},
for which a strong correlation is found (correlation coefficient $\sim 0.71$, 
Kendall's $\tau$ = 0.52).
This time the DNC lines show a systematical smaller broadening not only
in the HMSC group but also in the groups containing the more evolved
objects. In this case, however, the reason could be also attributed to the fact
that the two transitions require different excitation conditions: 
the (1--0) line traces gas colder, and hence more quiescent, 
than that associated with the emission of the (2--1) line.

\begin{figure*}
\begin{minipage}{160mm}
 \begin{center}
 \resizebox{\hsize}{!}{\includegraphics[angle=-90]{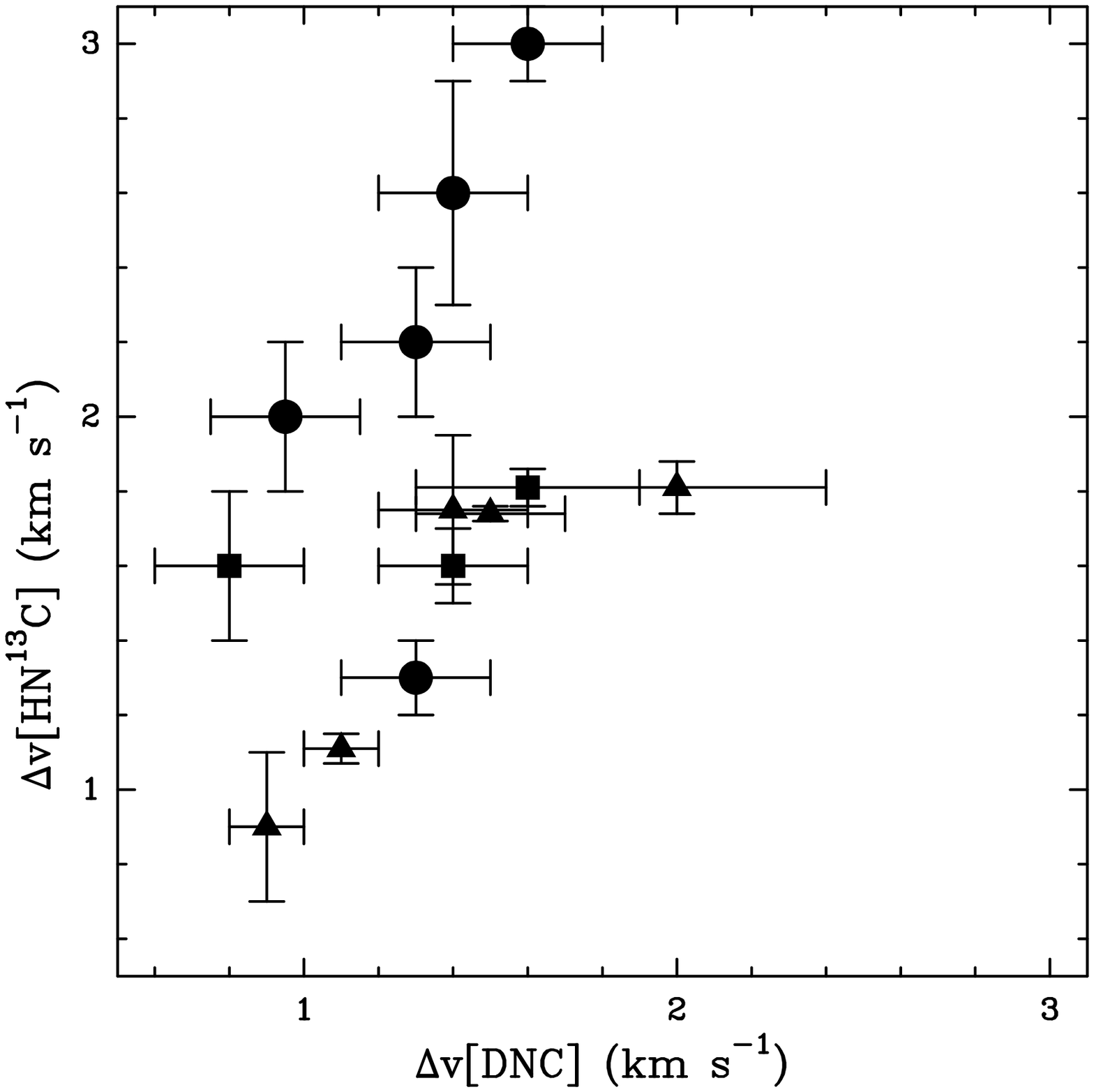}
                                      \includegraphics[angle=-90]{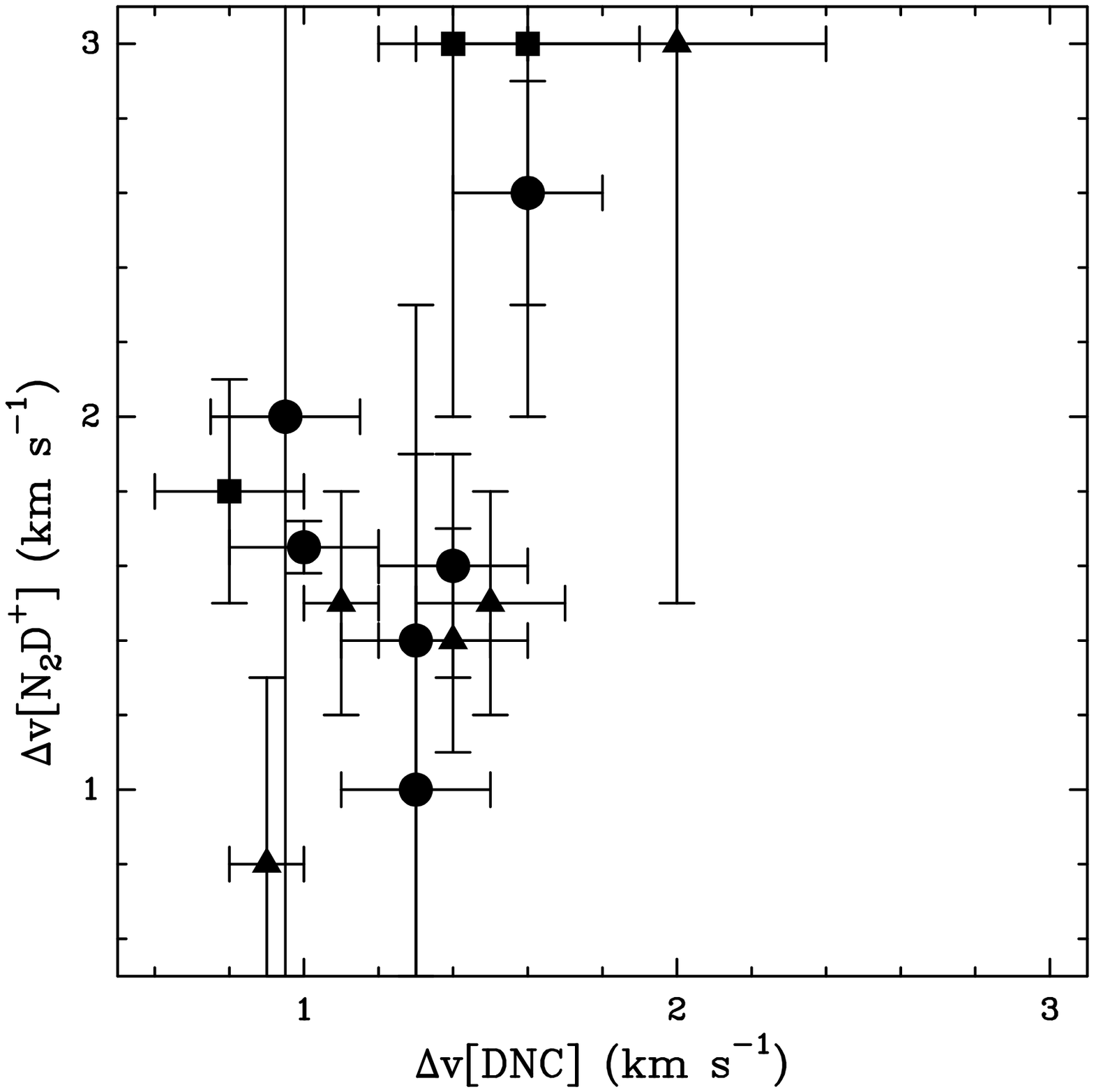}}
 \caption[]
 {\label{line_widths}{{\it Left panel:} Comparison between the line
 widths (full width at half maximum derived from Gaussian fits) for DNC(1--0) 
and \HNC (1--0). Only the sources detected in both lines are shown.
Circles indicate HMSCs, squares HMPOs and triangles UC HII regions. 
\newline
 {\it Right panel:} Same as left panel but the comparison is between
DNC(1--0) and \D (2--1).}}
 \end{center}
 \end{minipage}
 \end{figure*}

\subsection{Deuterated fraction of HNC}
\label{dfrac}

\begin{table}
\begin{center}
\caption[] {Best Gaussian fit line parameters for DNC(1--0). Between
parentheses, the errors calculated by the fitting procedure are given.}
\label{tab_fit_dnc}
\begin{tabular}{lcccc}
\hline
core & $A$ & $\Delta v$  & $T_{\rm pk}^{(1)}$ & rms$^{(2)}$ \\
         &  K \kms                     &   \kms\           &    K                  & K \\
\cline{1-5}
\multicolumn{5}{c}{HMSC} \\
\cline{1-5}
I00117--MM2 & 0.20(0.04) &	  1.0(0.2)	&     0.20(0.04) &    0.044 \\
G034--F2(MM7) &  0.38(0.06) &    1.3(0.2)  &     0.27(0.05) &    0.055  \\
G034--F1(MM8)  &  0.23(0.04) &     0.95(0.2) &	     0.23(0.04) &    0.017 \\
G034--G2(MM2)  & 0.53(0.06)  &   1.4(0.2)  &   0.35(0.05)  &    0.018 \\
G028--C1(MM9)  &  0.58(0.06) &   1.6(0.2)  &     0.35(0.05) &    0.032 \\
I20293--WC &    0.47(0.07) &	   1.3(0.2)	&     0.34(0.05) &    0.021 \\
I22134--B    & $\leq 0.14$  & & $\leq 0.04$ & \\
I22134--G    &  $\leq 0.14$  & & $\leq 0.035$ & \\
\cline{1-5}
\multicolumn{5}{c}{HMPO} \\
\cline{1-5}
I00117--MM1 & $\leq 0.19$ & & $\leq 0.04$ & \\
18089--1732 & 0.43(0.07) & 1.6(0.3) & 0.26(0.04) & 0.055 \\
18517+0437 &     0.43(0.05) &   1.4(0.2)	 &    0.29(0.05) &    0.012 \\
G75--core   &  $\leq 0.19$ &  & $\leq 0.05$  &  \\
 I20293--MM1 &  0.21(0.04) & 0.8(0.2) & 0.25(0.04) &  0.05 \\
 I21307             & $\leq 0.17$ & & $\leq 0.035$ & \\
 I23385            & $\leq 0.19$ & & $\leq 0.04$ & \\
\cline{1-5}
\multicolumn{5}{c}{UC HII} \\
\cline{1-5}
G5.89--0.39    &  0.72(0.07) &	  1.5(0.2)	 &    0.46(0.06) &   0.041 \\
I19035--VLA1 &   0.29(0.05) &     1.4(0.2)	  &   0.19(0.04)  &   0.018 \\
19410+2336  &  0.36(0.04)  &	   1.1(0.1)	 &     0.32(0.04) &    0.024 \\
ON1               &  0.40(0.06)    &    2.0(0.4)	&     0.18(0.04) &    0.026 \\
I22134-VLA1 & $\leq 0.16$ & & $\leq 0.04$ & \\
23033+5951 &    0.39(0.04) &   0.9(0.1)  &   0.40(0.02) &    0.030 \\
NGC7538--IRS9 & $\leq 0.16$ & & $\leq 0.04$ & \\
\hline
\end{tabular}
\end{center}
$(1)$ in $T_{\rm MB}$ units;
$(2)$ $1\sigma$ rms noise in the spectrum.
\end{table}

\begin{table}
\begin{center}
\caption[] {Same as Table~\ref{tab_fit_dnc} for \HNC (1--0). }
\label{tab_fit_hnc}
\begin{tabular}{lcccccccc}
\hline
core & $A$ & $\Delta v$  & $T_{\rm pk}^{(1)}$ & rms$^{(2)}$ \\
         &  K \kms                     &   \kms\           &    K                  & K \\
\cline{1-5}
\multicolumn{5}{c}{HMSC} \\
\cline{1-5}
 I00117--MM2 & $\leq 0.14$ & & $\leq 0.03$ & \\
G034--F2(MM7) &    1.05(0.06) & 	2.2(0.2)	&    0.44(0.04) &    0.05 \\
G034--F1(MM8) &   0.52(0.04)  &	 2.0(0.2)	&    0.25(0.03) &    0.04 \\
G034--G2(MM2) &    0.87(0.08)  &      2.6(0.3)  &      0.32(0.06) &    0.022 \\
G028--C1(MM9) &    1.69(0.08) &    	  3.0(0.1)	&      0.52(0.05) &   0.05 \\
I20293-WC         &   0.42(0.04)  &     	  1.3(0.1)	 &   0.30(0.04)  &    0.04 \\
I22134--B  & 0.09(0.02) & 0.7(0.2) & 0.36  & 0.03 \\
I22134--G &  0.35(0.02) &   0.93(0.07)  &   0.36(0.03)  &    0.03 \\
\cline{1-5}
\multicolumn{5}{c}{HMPO} \\
\cline{1-5}
I00117--MM1 & $\leq 0.14$ &   & $\leq 0.03$ & \\
18089--1732 &   1.96(0.05) & 	  1.81(0.05) &      1.02(0.04)  &    0.04 \\
18517+0437 &   0.79(0.04) &     1.6(0.1)    &     0.46(0.03)  &    0.03 \\
G75-core    &   0.43(0.04)    &     1.6(0.2)	 &    0.25(0.03)  &    0.03 \\
I20293--MM1  &   0.52(0.03)  &    1.22(0.09) &    0.40(0.03) &    0.03 \\
I21307       & $\leq 0.13$ &    &  $\leq 0.025$ & \\
I23385          &  0.15(0.04)    & 0.9(0.3)      & 0.08 & 0.03 \\
\cline{1-5}
\multicolumn{5}{c}{UC HII} \\
\cline{1-5}
G5.89--0.39  &     4.64(0.05) &   1.74(0.02) &      2.50	(0.04) &    0.03 \\    
I19035--VLA1 &  0.54(0.04)  &   1.75(0.2)  &	 0.29(0.03) &    0.03 \\
19410+2336  &  0.90(0.03)  &    1.11(0.04) &     0.76(0.03) &    0.03 \\
ON1              & 1.37(0.04)     &    1.81(0.07)  &     0.71(0.03)  &   0.03 \\
I22134-VLA1 &    0.18(0.03) &    0.9(0.2)   &	  0.19(0.03) &   0.02 \\
23033+5951 &  0.54(0.03)  & 	  1.2(0.1)  &        0.41(0.03)  &  0.02 \\
NGC7538--IRS9 &  0.57(0.04)  &  1.4(0.1) &	    0.39(0.03) &    0.03 \\
\hline
\end{tabular}
\end{center}
$(1)$ in $T_{\rm MB}$ units;
$(2)$ $1\sigma$ rms noise in the spectrum.
\end{table}

To derive \Dfrac (HNC) from the line parameters, we have adopted
the same approach as in Sakai et al.~(\citeyear{sakai12}). First, we assume the lines
are optically thin. This is justified by both the shape of the lines (which have not
the flat-topped shape typical of high optical depth transitions) and 
by the findings of Sakai et al.~(\citeyear{sakai12}) in similar objects. Second,
given the similar critical density of the two transitions ($\sim 0.5\times 10^6$ \cmc ), 
we assume their {excitation temperatures} and emitting regions, and thus 
the filling factors, are the same. 
Under these hypotheses, the column density ratio is given by (see Sakai et
al.~\citeyear{sakai12}):
\begin{equation}
\frac{N({\rm DNC})}{N({\rm HN^{13}C})} \simeq 1.30{\rm exp}\left(-\frac{0.52}{T_{\rm ex}}\right)\frac{A_{\rm DNC}}{A_{\rm HN^{13}C}}
\label{eq_dfrac}
\end{equation}
where 
\Tex\ is the excitation temperature,
and $A$ is the integrated area of the line (in $T_{\rm MB}$ units).

The method adopted to fit the lines does not allow to derive directly
\Tex , for which a good fit to the hyperfine structure is needed. Therefore, 
in Eq.~(\ref{eq_dfrac}) as \Tex\ we have taken the kinetic temperatures 
given in Fontani et al.~(\citeyear{fontani11}, see their Table~3)
assuming that the lines are thermalised.
These have been derived either directly from the \H\ (3--2) transition 
or from \AMM\ measurements, and then extrapolated to \Tk\ following 
Tafalla et al.~(\citeyear{tafalla}). In principle, the excitation temperature
derived from \H\ or \AMM\ can be different from that of the DNC and \HNC\
(1--0) lines, but we stress that
Eq.~(\ref{eq_dfrac}) shows that the column density ratio is
little sensitive even to changes of an order of magnitude in \Tex\
(see also Sakai et al.~\citeyear{sakai12}).

We have then converted $N({\rm DNC})/N({\rm HN^{13}C})$ into 
$N({\rm DNC})/N({\rm HN^{12}C})$=\Dfrac (HNC) by 
calculating the $^{13}$C/$^{12}$C abundance ratio from the relation: 
$^{13}$C/$^{12}$C$ = 1/(7.5\times D_{\rm gc} + 7.6)$ (Wilson \& Rood~\citeyear{wer}),
where $D_{\rm gc}$ is the source Galactocentric distance in kpc, and multiplying 
Eq.~(\ref{eq_dfrac}) by this correction factor.
In Table~\ref{tab_dfrac} we list the column density ratio \Dfrac (HNC)
for the cores observed and the physical parameters used to derive it
as explained above: Galactic coordinates (longitude $l$, latitude $b$), Galactocentric
distance ($D_{\rm gc}$), isotopic abundance ratio $^{13}$C/$^{12}$C,
$A_{\rm DNC}$, $A_{\rm HN^{13}C}$, and $T_{\rm ex}$.

\begin{table*}
\begin{center}
\caption[] {}
\label{tab_dfrac}
\begin{tabular}{lcccccccc}
\hline
 core &          $ l$ &  $ b $  &   $D_{\rm gc}$  &  $^{13}$C/$^{12}$C & $A_{\rm DNC}$  & $A_{\rm HN^{13}C}$  &    $T_{\rm ex}^{(1)}$ & \Dfrac (HNC) \\
         &         $^o$  & $^o$  &  kpc     &       &    K \kms\ & K \kms & K & \\
\cline{1-9}
\multicolumn{9}{c}{HMPO} \\
\cline{1-9}
 I00117--MM2    &   2.082  &   0.03706      &      9.5  &  79  &      0.20	&    $\leq 0.14$ & 14    &  $\geq 0.008$  \\
 G034--F2(MM7)  &    0.6071  & --0.005093 &      5.9   &  52 &     0.38	        &    1.05   & 17 &    0.009 \\
 G034--F1(MM8)   &   0.6068  & --0.005021   &    5.9   &  52   &    0.23	&    0.52   &  17 &  0.011 \\
 G034--G2(MM2)   &   0.6132  &   --0.0192    &     6.4 &  55   &   0.53   	&   0.87   &  17 &  0.014 \\
 G028--C1(MM9)  &    0.5004 &  --0.008629  &    4.8    &  43   &	0.58	  &    1.69   &  17 & 0.010 \\
I20293--WC    &   1.384  &  0.003144      & 8.4     &  70  &   0.47 	&   0.42  & 17   &  0.02  \\
I22134--B     &  1.819   &  0.03383    &   9.5     &  79  &  $\leq 0.14$     &  0.09   & 17    &   $\leq 0.025$ \\ 
I22134--G     &  1.819  &   0.03373  &     9.5     &  79  &  $\leq 0.14$    &  0.35   & 25  &    $\leq 0.006$  \\
 \cline{1-9}
\multicolumn{9}{c}{HMSC} \\
\cline{1-9}
18089--1732   &     0.231  & --0.001963   &   5.1    &  46  &  0.43   &    1.96    &  38 &   0.006  \\
18517+0437  &    0.6592 &    0.01745     &    6.5   &  56   &  0.43   &   0.79   & 43 &    0.0125  \\
 G75-core      & 1.329  &  0.002301       &    8.4    &  71  &  $\leq  0.19$   &    0.43    &  96 &  $\leq 0.008$  \\
I20293-MM1  &     1.385 &   0.003035     &   8.4    &  70  &    0.21   &    0.52    &  43 &   0.008  \\
I23385        &   2.005   & --0.006124   &    11.5   &  94   &   $\leq 0.19$     &    0.15  &  43 &  $\leq 0.017$ \\ 
\cline{1-9}
\multicolumn{9}{c}{UC HII} \\
\cline{1-9}
G5.89--0.39  &    0.1088 &   --0.01743    &    7.2   &  62  &    0.72   &    4.64   &  26 &   0.003  \\
I19035--VLA1  &    0.7151  &  --0.01113   &    7.0    &  60  &    0.29   &     0.54  & 39 &   0.011 \\ 
19410+2336   &     1.05 &  --0.005147   &    7.7    &  65 &     0.36   &   0.90   &  21 &   0.008  \\
 ON1                    &    1.22  &  --0.02177   &    8.0   &  68  &    0.40 	&   1.37 & 26   &    0.006 \\  
 I22134--VLA1   &    1.819 &     0.0338    &   9.5     &  79  &     $\leq 0.16$   &     0.18 &  47    &  $\leq  0.014$  \\
  23033+5951    &   1.928  &  0.001401    &   10.3     &  85  &  0.39   &    0.54  &  25 &  0.011 \\ 
NGC7538-IRS9   &    1.953 &    0.01593   &    9.9     &  82  &  $\leq 0.16$    &    0.57  &  26  &  $\leq 0.004$ \\ 
 \hline
\end{tabular}
\end{center}
$^{(1)}$ assumed equal to the gas kinetic temperature listed in Table~A.3 of Fontani et al.~(\citeyear{fontani11}).
\end{table*}

\begin{figure}
\begin{minipage}{80mm}
 \begin{center}
 \resizebox{\hsize}{!}{\includegraphics[angle=-90,width=10cm]{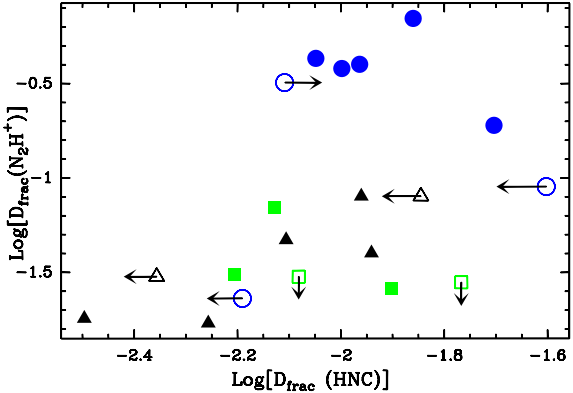}}
 \caption[]
 {\label{fig_comp}{Comparison between the deuterated fraction of \H ,
\Dfrac (\H ) =$N$(\D )/$N$(\H ), and that of HNC, \Dfrac (HNC)=$N$(DNC)/$N$(HNC), 
in the cores studied by Fontani et al.~(\citeyear{fontani11}). Only the sources detected
in DNC(1--0) or \HNC (1--0) are shown. Circles indicate HMSCs, squares HMPOs, 
and triangles UC HII regions. Empty symbols with arrows indicate lower or upper limits, 
depending on arrow orientation.}}
 \end{center}
 \end{minipage}
 \end{figure}
 
\section{Discussion}
\label{dis}

\subsection{\Dfrac (\H ) versus \Dfrac (HNC)}
\label{dfrac_comp}

Table~\ref{tab_dfrac} shows that the HMSC group has the highest average \Dfrac (HNC)
(mean value $\sim 0.012$, $\sim 0.019$ if one includes the lower
limit on I00117--MM2). The HMPOs and UC HII groups have very similar
average \Dfrac (HNC) ($\sim 0.009$ and $\sim 0.008$, respectively),
but given the dispersion and the poor statistics, there are no 
significant statistical differences between the three groups.
By comparing the observational data of this work and those of Fontani et al.~(\citeyear{fontani11}),
we clearly note a different behavior of \Dfrac (HNC) and \Dfrac (\H ) in high-mass star-forming
cores: \Dfrac (HNC) does not change significantly going from the pre-protostellar phase to
subsequent phases of active star formation, while \Dfrac (\H ) is smaller of an order of magnitude in the 
evolved phases (HMPOs and UCHIIs) than in the pre-protostellar phase (HMSCs), and
the latter evolutionary group is undoubtedly statistically separated from 
the other two (Fontani et al.~\citeyear{fontani11}). We stress
once more that these results, obtained towards the same clumps and with comparable telescope
beam sizes, are not affected by possible biases introduced by the source selection.

In Fig.~\ref{fig_comp} we compare \Dfrac (\H ) and \Dfrac (HNC) measured 
in the cores observed in both \D\ (Fontani et al.~\citeyear{fontani11}) and DNC.
As one can see, the sources with the largest  \Dfrac (\H ) tend to also have
the largest \Dfrac (HNC), despite the different absolute magnitude 
especially in the HMSC group. In fact, the two parameters
are slightly correlated, with a Kendall's $\tau$ rank correlation coefficient
of $\sim 0.36$ (excluding lower and upper limits, which tend to reinforce
the possible correlation though). 

\subsection{\Dfrac (\H ) and \Dfrac (HNC) versus chemical models}
\label{model}

The results presented in Section~\ref{dfrac_comp} are consistent overall with the 
scenario proposed by the chemical models of Sakai et al.~(\citeyear{sakai12}): 
the \D /\H\ abundance ratio sharply decreases after the protostellar birth, 
while the DNC/HNC abundance ratio decreases more gradually and maintains 
for longer the high deuteration of the earliest evolutionary stages of the core.
In this subsection, we compare the observational results with the model predictions 
in detail.

We have solved the chemical rate equations with the state-of-art gas-grain reaction network of Aikawa 
et al.~(\citeyear{aikawa12}).
The model includes gas-phase reactions, interaction between gas and grains, and grain surface reactions.
The parameters for chemical processes are essentially the same as in Sakai et al.~(\citeyear{sakai12}), 
except for the binding energy of HCN and HNC; we adopt a binding energy of 4170 K for HCN 
(Yamamoto et al.~\citeyear{yamamoto}), while a smaller value of 2050~K was used for those 
species in Sakai et al.~(\citeyear{sakai12}). We assume the binding energy of HNC to be the 
same as that of HCN. Species are initially assumed to be atoms (either neutrals or ions), except for 
hydrogen and deuterium, which are in molecular form. The elemental abundance of deuterium 
is set to be $1.5\times10^{-5}$ (Linsky 2003). As a physical model, we assume a static homogeneous 
cloud core with a H$_2$ volume density of 10$^4$ cm$^{-3}$ or 10$^5$ cm$^{-3}$. 
To mimic the protostar formation, we suddenly rise a temperature from 15 K to 40 K 
at a given time of 10$^5$ yr. The two temperatures are consistent with the average kinetic
temperatures measured in the targets of this work ($T=18$~K and $T=38$~K for HMSCs and
HMPOs, respectively, see Table~\ref{tab_dfrac}).
Our choice of 10$^5$ yr is comparable to the timescale of high-mass starless-phase 
($3.7\times10^5$ yr) estimated by the statistical study of Chambers et al.~(\citeyear{chambers09}), 
while Parsons et al. ~(\citeyear{parsons09}) estimated it to be a few 10$^3$--10$^4$ yr.

\begin{figure*}
\begin{minipage}{180mm}
 \begin{center}
 \resizebox{\hsize}{!}{\includegraphics[angle=0]{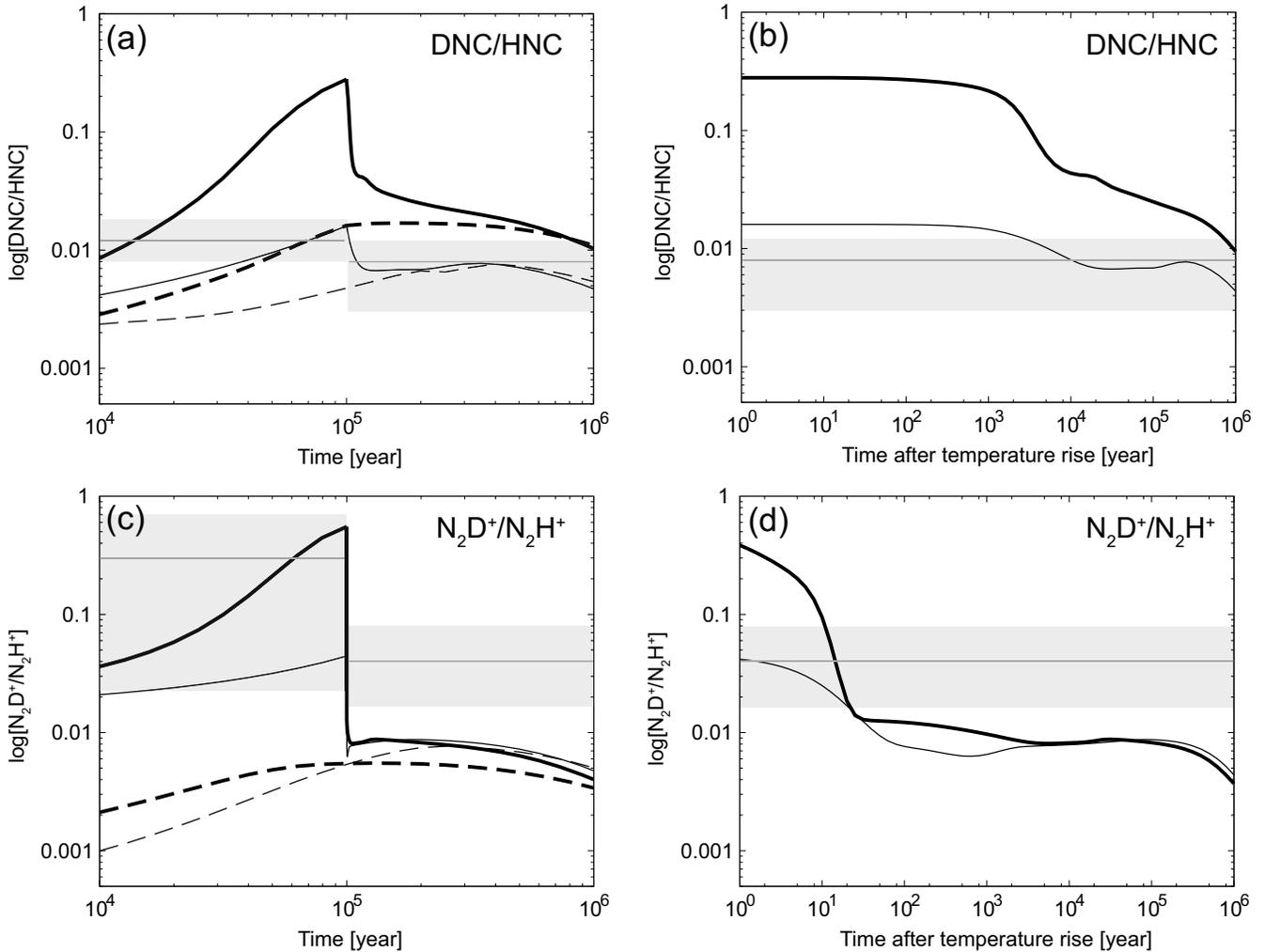}
                                      }
 \caption[]
 {\label{models_data}{{\bf (a):} temporal variation of the DNC/HNC abundance ratio as predicted
 by the chemical model described in Sect.~\ref{model}. 
 Solid lines represent the model with a temperature jump from 15 to 40~K at $10^5$ yrs, 
 while dashed lines represent the model with constant temperature of 40 K during 10$^6$ yrs. 
 Thin and thick lines represent the model with $n_{\rm H_2} = 10^4$ cm$^{-3}$ and $10^5$ cm$^{-3}$, 
 respectively.
 The horizontal grey lines and areas indicate the average values and dispersion obtained from the
 observational data in this work.
 \newline
{\bf (b):} same as {\bf (a)} for \D /\H . The observational data here are from Fontani et al.~(\citeyear{fontani11}). 
 \newline
{\bf (c)} and {\bf (d):} same as {\bf (a)} and {\bf (b)}, respectively, but the time starts from when the temperature increases 
from 15 K to 40 K.}}
 \end{center}
 \end{minipage}
 \end{figure*}

A comparison between model predictions and observational results is illustrated in 
Fig.~\ref{models_data}. In our model, in the pre-protostellar stage ($T=15$ K) 
the deuteration of both molecules increases similarly with time: this is due to the fact that
the deuterium fractionation is initiated in both species by the same
route reaction (i.e. H$_3^+$ isotopologues), as mentioned in Sect.~\ref{intro}. 
On the other hand, if we assume that the
deuterated fractions of the HMSCs must be compared with the model predictions
before the temperature rise, i.e. before $t\simeq10^5$ yr,
the measured \Dfrac (HNC) and \Dfrac (\H ) cannot
be reproduced simultaneously in a single model: \Dfrac (HNC) can be well 
reproduced with $n_{\rm H_2} = 10^4$ cm$^{-3}$, while \Dfrac (\H ) is 
reproduced with $n_{\rm H_2} = 10^5$ cm$^{-3}$.
This discrepancy could indicate that the observed lines of DNC 
(and HN$^{13}$C) arise from regions that, on average, 
are less dense than those responsible for the emission of \D\ (and \H). 
In fact, the observational parameters are averaged values measured over 
slightly different angular regions ($\sim 21$\asec\ for DNC, and $\sim 16$\asec\
for \D ), so that the emission seen in DNC could be more affected
than that seen in \D\ by the contribution from the low-density envelope 
surrounding the dense cores, where the deuterium fractionation is expected to
be less important. This can explain why the model 
with lower average gas density can reproduce \Dfrac (HNC), 
but not \Dfrac (\H ), for which a higher average gas density is needed.
Another possibility is that we are missing something in the current chemical model.
For example, we do not consider the ortho state of hydrogen molecules in this work.
The presence of ortho-H$_2$ suppresses the deuteration process of molecules, 
since the internal energy of ortho-H$_2$ helps to 
overcome the endothermicity of reaction (1) in the backward direction (Flower et al.~\citeyear{flower06}).
If we consider ortho-H$_2$, however, both the DNC/HNC and \D/\H\ abundance ratios would be lowered.

Inspection of Fig.~\ref{models_data} also shows that an average 
\Dfrac (\H )$\geq 0.2$ in a starless core with $n_{\rm H_2} = 10^5$ cm$^{-3}$ 
is reached at a time close to $\sim 10^5$ yrs. Assuming this as the time necessary 
for the starless core to collapse, as suggested by Chambers et al.~(\citeyear{chambers09}), 
this means that only cores relatively 
close to the onset of gravitational collapse, i.e. the so--called pre--stellar cores, 
can give rise to the observed high values of \Dfrac (\H ). This behaviour is 
in agreement with the predictions of chemical models including also the spin 
states of the H$_2$ and H$_3^+$ isotopologues (Kong et al.~\citeyear{kong}),
in which levels of \Dfrac (\H ) larger than 0.1 are possible only in cores older
than $\sim 10^5$ yrs. Models with higher average
density ($\geq 10^6$\cmc ) can reproduce such high deuterated fractions in shorter 
times, but these average densities are not realistic to represent regions with angular 
sizes of $\sim 16$\asec , like those that we have observed. 

In the protostellar stage ($T=40$ K), the \D/\H\ abundance ratio sharply decreases 
in timescale of $\sim$10$^2$ yrs, while the DNC/HNC abundance ratio decreases in 
$\sim$10$^4$ yrs both in the model with $n_{\rm H_2} = 10^4$ cm$^{-3}$ and in that
with $10^5$ cm$^{-3}$.
Again, the measured \Dfrac (HNC) is better reproduced by the model with 
$n_{\rm H_2} = 10^4$ cm$^{-3}$ after 10$^4$--10$^5$ yr from a temperature rise,
while our model slightly underestimates \Dfrac (\H ), regardless of the
average gas density, unless the protostellar cores are extremely young
(i.e. age shorter than $10^2$ yrs).
It should be noted that once the gas temperature rises, not only the \D /\H\ ratio,
but also the \H\ abundance decreases significantly in the models. Also, according
to models of spherical star-forming cores (Aikawa et al.~\citeyear{aikawa12},
Lee et al.~\citeyear{lee}), the central warm gas is still surrounded by a spherical
shell of cold and dense gas during the early stages of collapse, in which
both \D\ and \H\ are still abundant. This is not taken into account for simplicity
in our one-box model. However, such residual emission from the cold envelope could
explain the higher \D /\H\ ratio still apparent after the sudden temperature rise.

\section{Conclusions}
\label{con}

We have observed the DNC and \HNC (1--0) rotational transitions
towards 22 massive star-forming cores in different evolutionary
stages, towards which \Dfrac (\H) was already measured
by Fontani et al.~(\citeyear{fontani11}). The aim of the work
was to compare \Dfrac (HNC) to \Dfrac (\H ) in the same sample of sources 
and with similar telescope beams, so that the comparison should not suffer 
from possible inconsistencies due to different sample selection criteria. 
The main observational result of this work confirms the predictions of the models
of Sakai et al.~(\citeyear{sakai12}), namely that
\Dfrac (HNC) is less sensitive than \Dfrac (\H ) to a sudden temperature rise,
and hence it should keep more than \Dfrac (\H ) of the thermal history of
the cores, despite the chemical processes leading to the
deuteration of the two species are similar. 
Therefore, our work clearly indicates that \Dfrac (\H ) is more suitable than
\Dfrac (HNC) to identify high-mass starless cores.
Based on the predictions of our chemical models,
the starless cores studied in this work having \Dfrac (\H ) around 0.2 - 0.3 
are very good candidate massive 'pre--stellar' cores, because only relatively
'evolved' starless cores can be associated with such high values of \Dfrac (\H ). 
Several results require follow-up high-angular resolution observations
to map the emitting region of DNC and \D , as well as DNC and \HNC ,
and test if these are slightly different as the present low-angular resolution 
data seem to suggest. 
Observations of higher excitation \HNC\ and DNC lines (3--2 or 4--3) may 
be also important to constrain better the excitation conditions.

\section*{Acknowledgments}

We thank the NRO staff for their help in the observations with the
45m Telescope presented in this paper. The 45m Telescope is
operated by the Nobeyama Radio Observatory, a branch of the
National Astronomical Observatory of Japan. This study is 
supported by KAKENHI (21224002, 23540266,
25400225 and 25108005).
K.F. is supported by the Research Fellowship from the Japan 
Society for the Promotion of Science (JSPS) for Young Scientists.

{}

\newpage

\renewcommand{\thefigure}{A-\arabic{figure}}
\setcounter{figure}{0}
\section*{Appendix A: Spectra}
\label{appb}

In this appendix we show all spectra of the DNC(1--0) and \HNC (1--0) transitions.

\begin{figure*}
\centerline{
                     \includegraphics[angle=0,width=16cm]{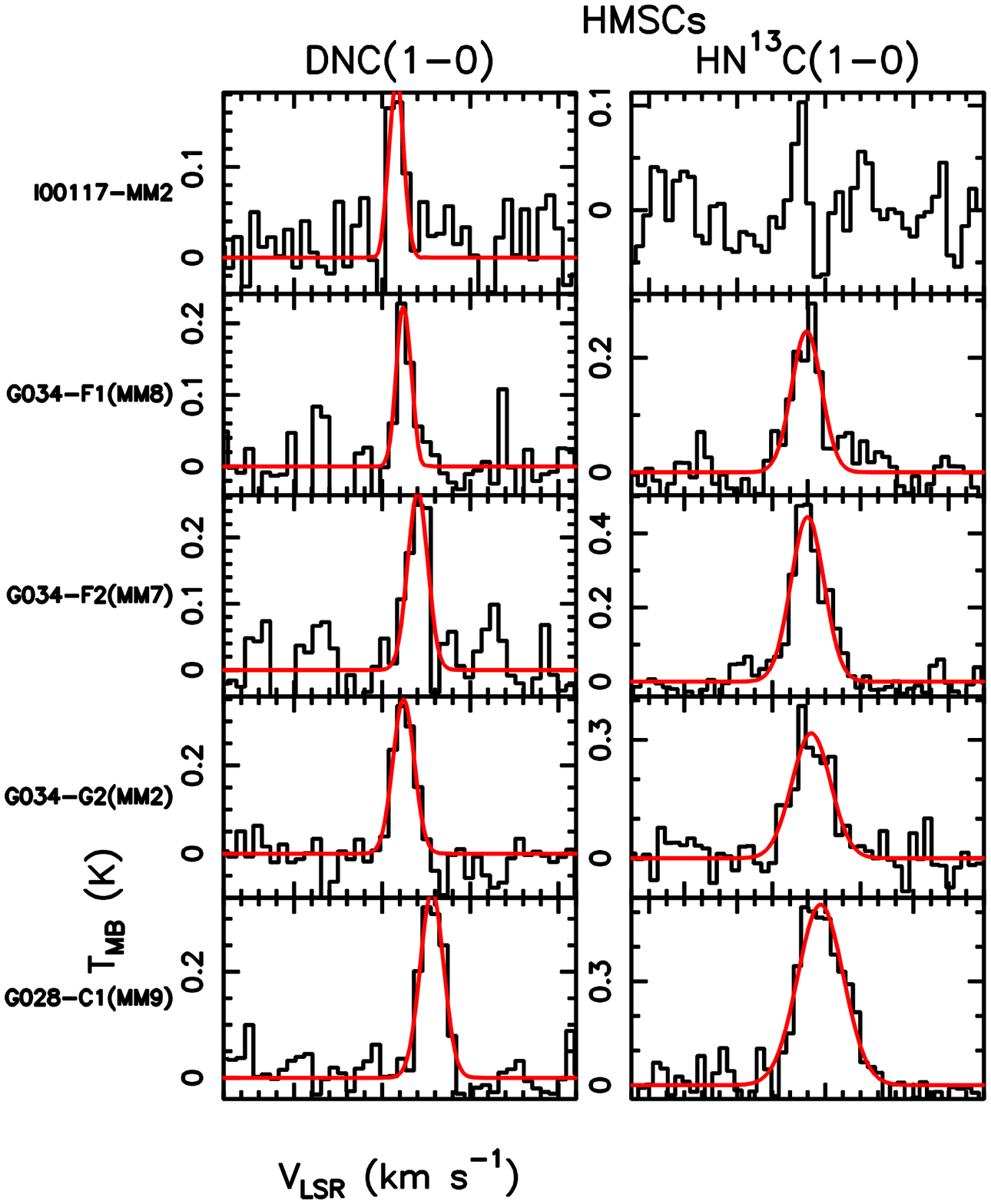}}
\caption[]{Spectra of DNC(1--0) and \HNC (1--0) obtained towards the sources 
classified as HMSCs. 
For each spectrum, the velocity interval shown is $\pm 10$ \kms\ from the
systemic velocity listed in Table~\ref{tab_sou}.
The y-axis is in main beam brightness temperature units. 
In each spectrum the red curve represents the best Gaussian
fit to the lines, the parameters of which are listed in Tables~3 and 4.}
\label{spectra1}
\end{figure*}

\begin{figure*}
\centerline{
                     \includegraphics[angle=0,width=16cm]{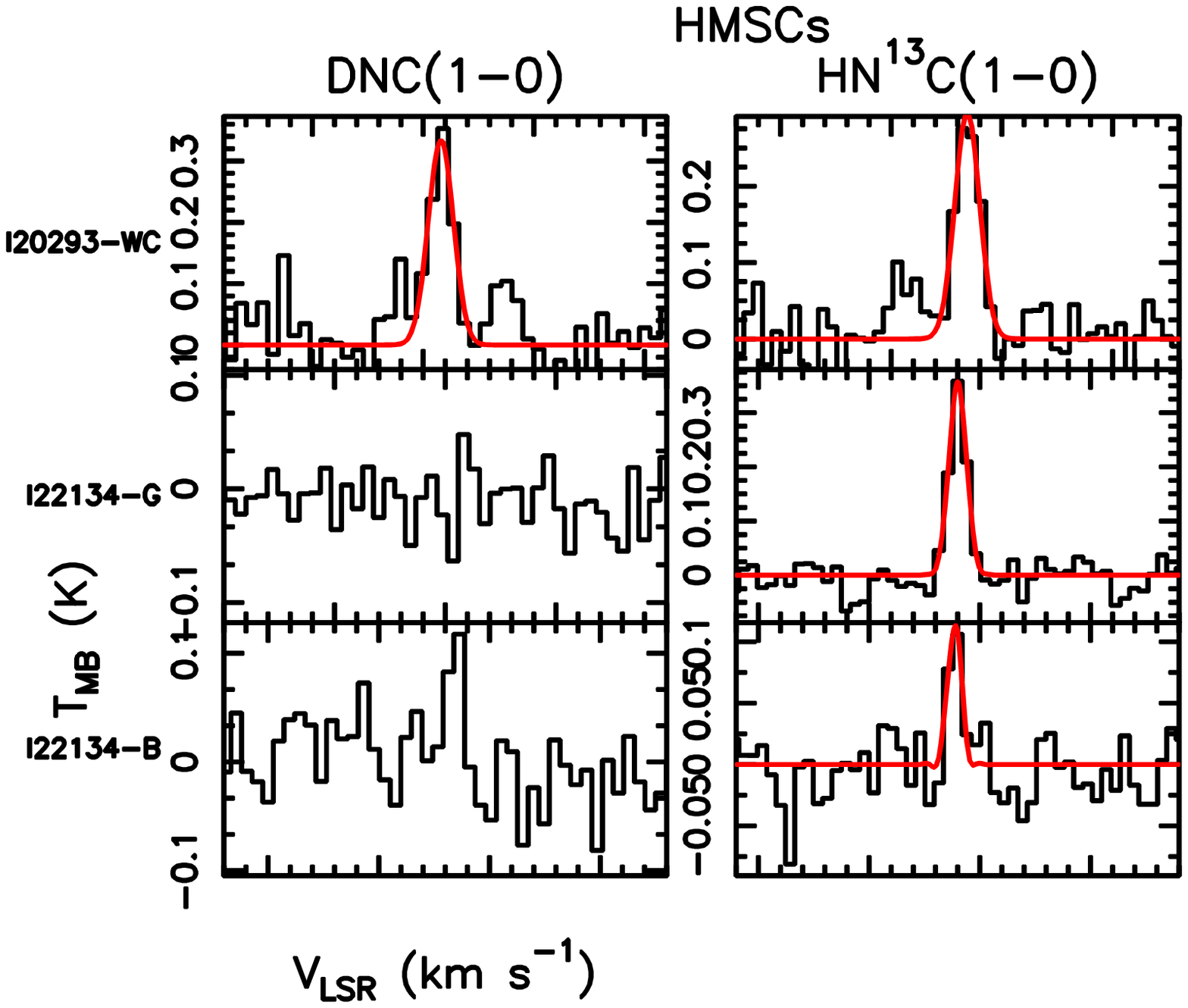}}
\caption[]{Fig.~A.1 continued. }
\label{spectra2}
\end{figure*}

\begin{figure*}
\centerline{
                     \includegraphics[angle=0,width=16cm]{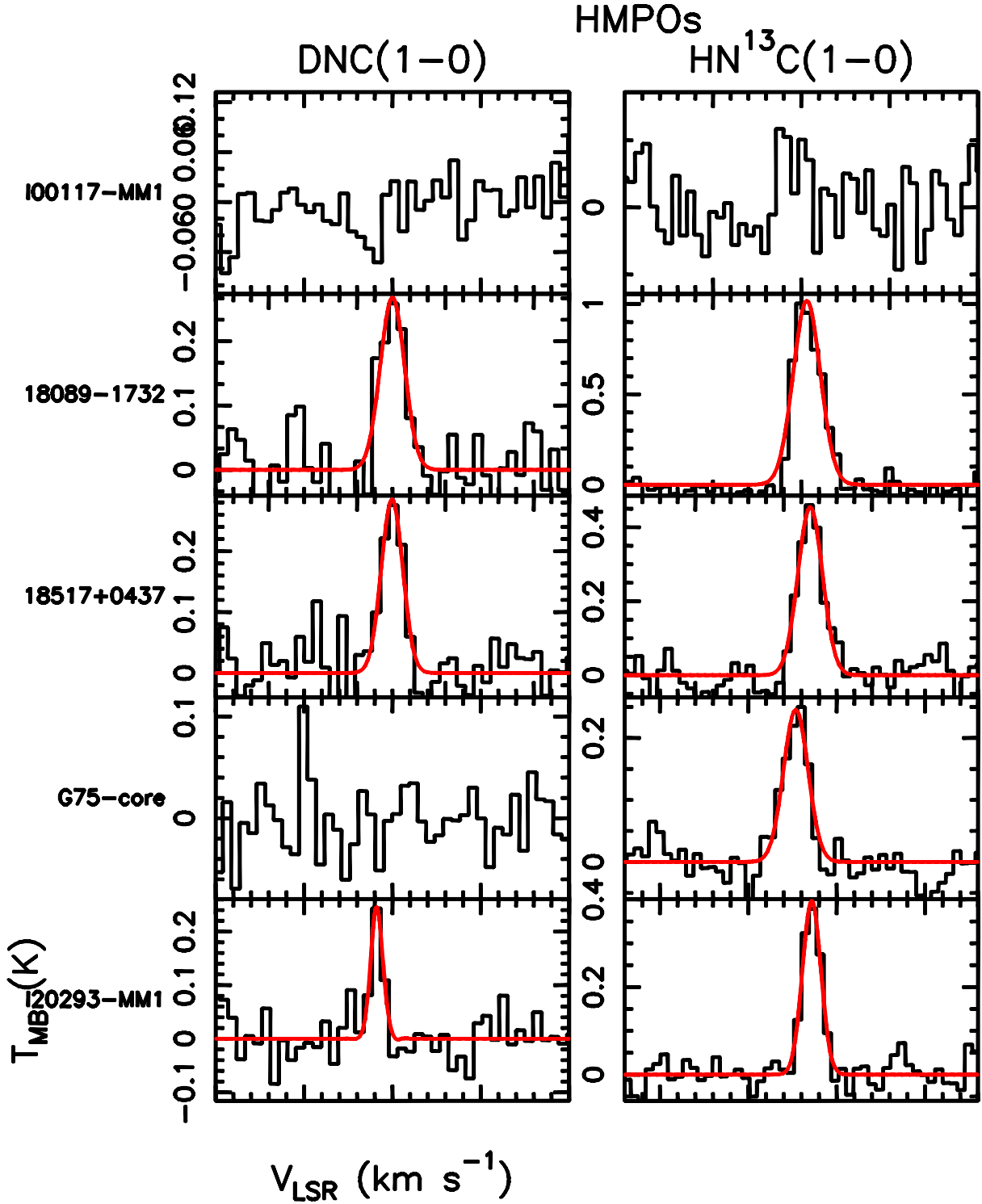}}
\caption[]{Same as Fig~A.1 for HMPOs. }
\label{spectra3}
\end{figure*}

\begin{figure*}
\centerline{
                     \includegraphics[angle=0,width=16cm]{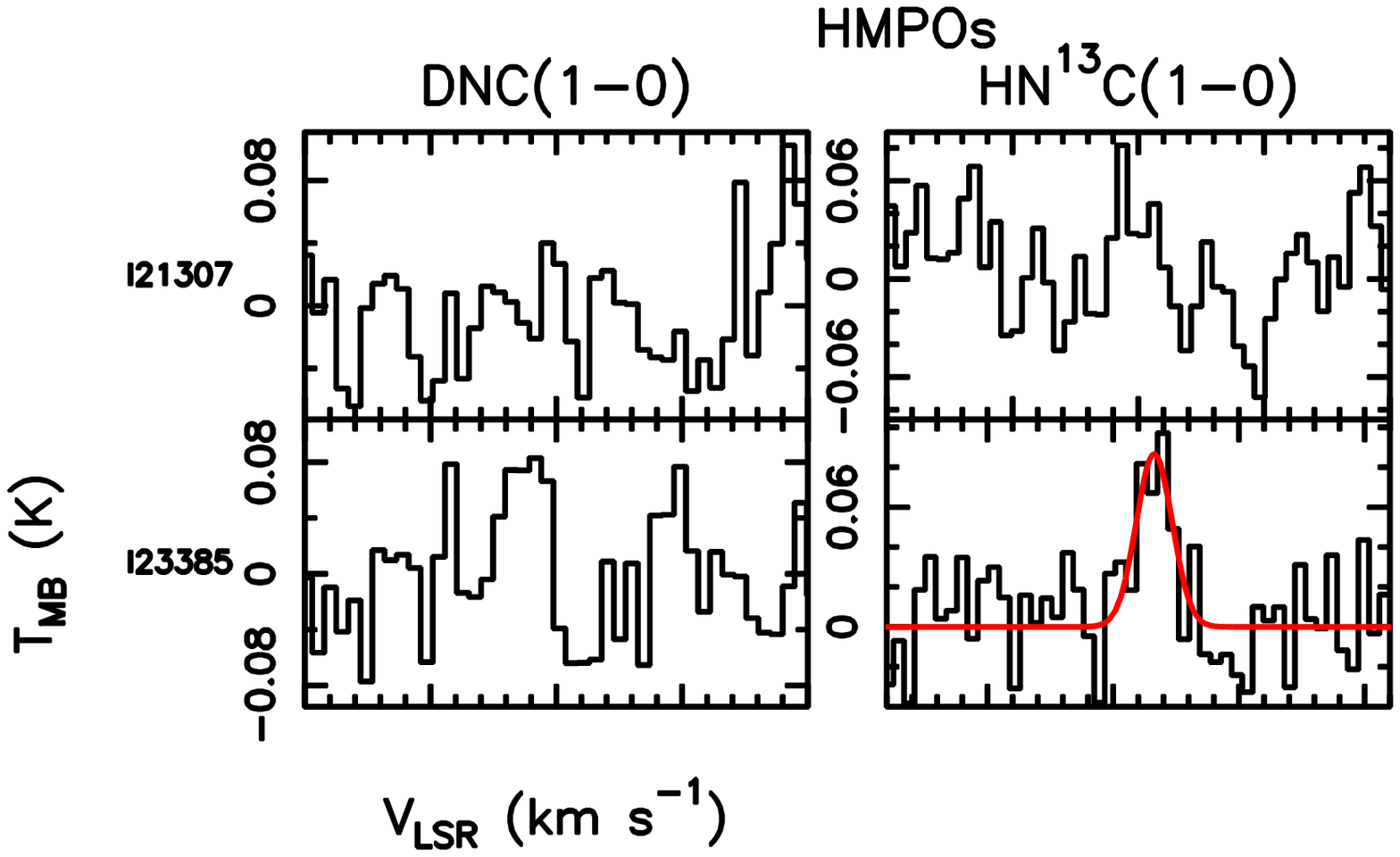}}
\caption[]{Fig.~A.3 continued. }
\label{spectra4}
\end{figure*}

\begin{figure*}
\centerline{
                     \includegraphics[angle=0,width=16cm]{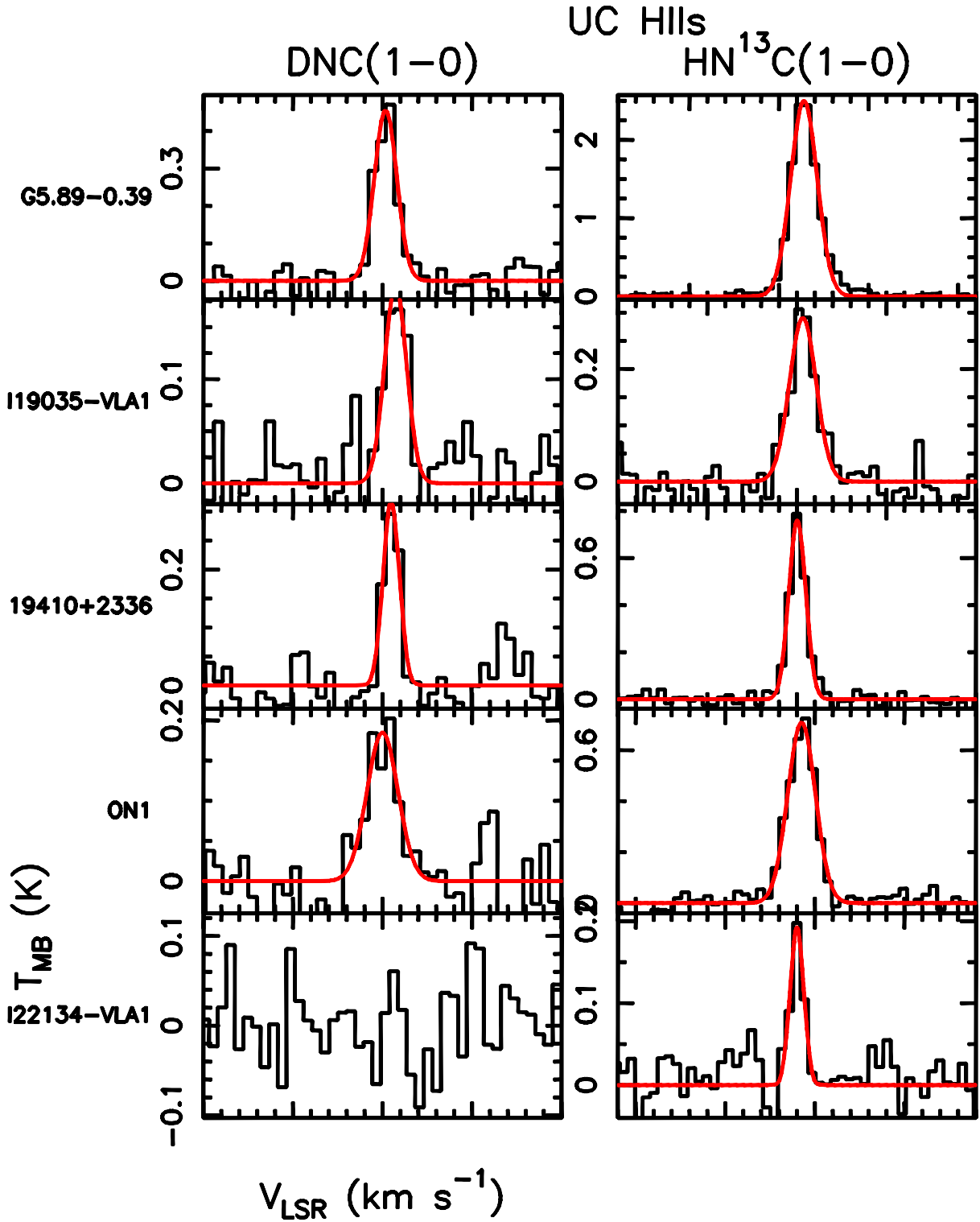}}
\caption[]{Same as Fig.~A.1 for UC HII regions. }
\label{spectra5}
\end{figure*}

\begin{figure*}
\centerline{
                     \includegraphics[angle=0,width=16cm]{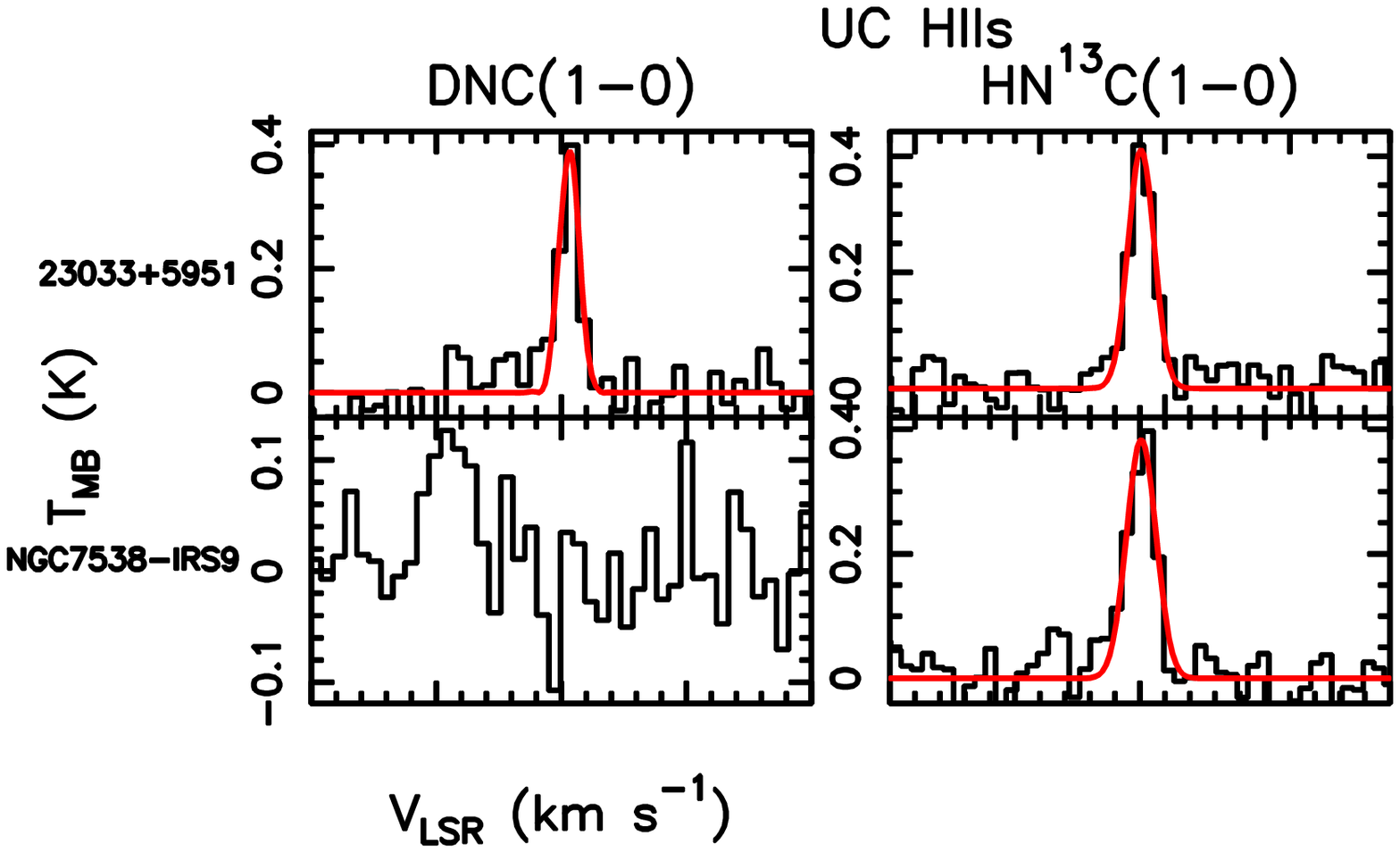}}
\caption[]{Fig.~A.5 continued. }
\label{spectra6}
\end{figure*}

\end{document}